\documentstyle[11pt,paspconf,epsf]{article}

\begin{document}

\title{ Spectroscopy of red giants of the Sagittarius dwarf galaxy}
\author{
G. Marconi}
\affil{Osservatorio Astronomico di Roma}
\author{
P. Bonifacio}
\affil{Osservatorio Astronomico di Trieste}
\author{L. Pasquini}
\affil{European Southern Observatory}
\author{P. Molaro}
\affil{Osservatorio Astronomico di Trieste}

\begin{abstract}
The photometric study of the Sagittarius dwarf galaxy 
by Marconi et al (1998) has suggested 
the presence of a
spread in metallicity (-0.7$\le$ [Fe/H] $\le$ -1.6) , which 
may result from  different bursts of star formation.
We present the results from 
a Multi Object Slit spectroscopy program carried out
at the NTT with the aim
to determine 
spectroscopic abundances of Sagittarius giants selected from the CCD
photometry. From our intermediate resolution
(R$\sim$1500 ) spectra radial velocities are determined to confirm the
membership and metallicities are derived by using spectral synthesis 
codes for stars
down to V$\sim$18, i.e. below the RGB clump. 
Out of 57 observed stars 23 have a radial velocity consistent
with Sagittarius membership, here we present results for 8 of these.
No star with  [Fe/H]$<-1.0$ is found, three stars are found to be
metal-rich. 
\end{abstract}


\section{Introduction}

In a study of the kinematics and abundances 
of the outer regions of the Galactic Bulge
Ibata, Gilmore \& Irwin (1994, 1995) discovered
a group of stars with heliocentric radial
velocity around 140 kms$^{-1}$ and a low velocity dispersion,
which could not be accounted for by any model of the Galactic
Bulge. The finding was readily interpreted as
a dwarf spheroidal galaxy (dSph) which is
a satellite of our own Galaxy.
At a distance of about 25 kpc Sagittarius is the nearest external
galaxy. It is undergoing
tidal disruption and is probably in the process
of merging with the  Galaxy.    
This could in fact be one of
the   many merger events which have contributed to shape
the present day Galaxy.
The colour-magnitude diagram of Sagittarius  suggests
a spread in  metallicity,  which may  result from  different
bursts of star formation.
We have analyzed low resolution spectra NTT+EMMI/MOS
to obtain radial velocities and crude
estimates of abundances.
The method we have developed is quite powerful
but the low S/N obtainable with NTT on these V$\approx 18.5$
stars  does not allow any firm conclusion on the chemical
history of Sagittarius.

\section{Observations}

\begin{figure}[t]
\plotone{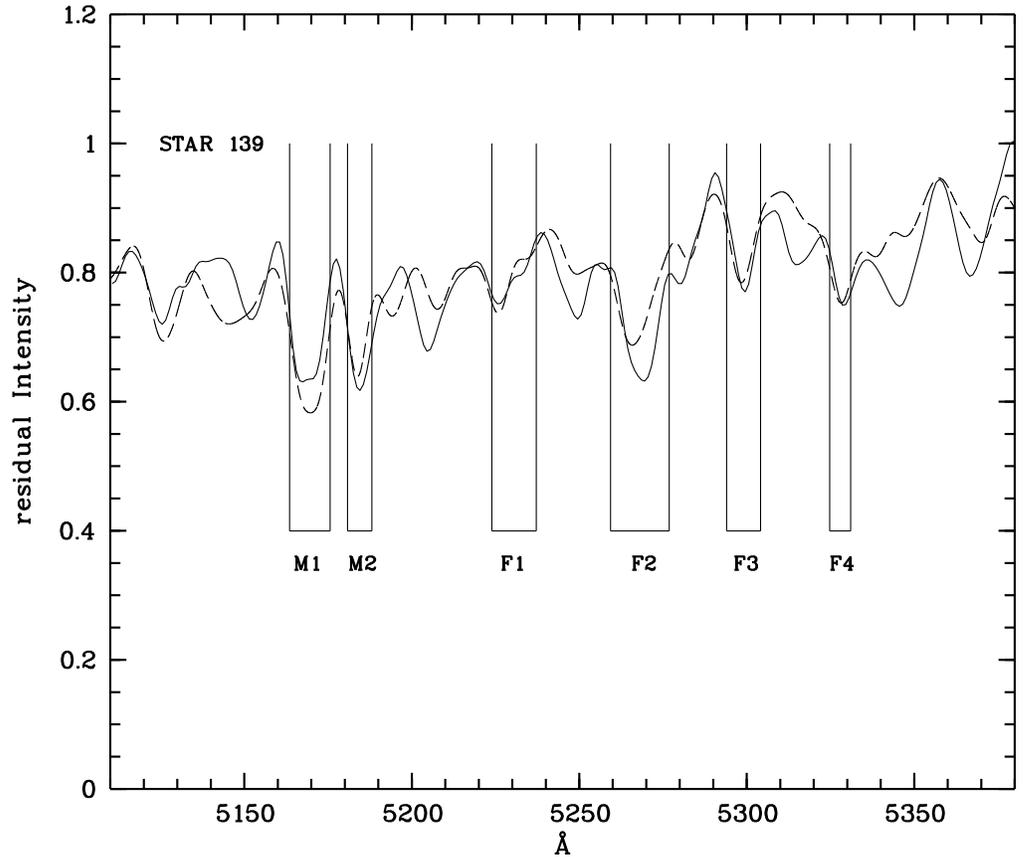}
\caption{Observed (solid line) and computed spectrum (dashed line)
for star 139. The spectra have been smoothed with a gaussian of
width 300 kms$^{-1}$ for display purposes.
The solid boxes show the spectrophotometric indices used. }
\end{figure}
The observations described here  were taken
at ESO La Silla on June 19th 1996 with the NTT telescope
and EMMI in multi-object-spectroscopy (MOS) mode.
We used grism \#5, which provides a resolving power of
about $R=\lambda/\Delta\lambda \sim 1500$. 
The range covered varies slightly depending on the
position of the star relative to the center of the field,
but the range 480-550nm was available for all the stars.
The spectra are not useful
towards shorter wavelengths due to the falling stellar flux
and the decreased efficiency of the CCD.
The log of observations
is given in table 1. Details on the observed field
can be found in Marconi et al (1998).

\begin{table}[t]
\caption{Observations log}
\begin{center}
\begin{tabular}{cccc}
\tableline
date &\# of slits & field & $T_{exp}(s) $\\
\tableline
1998/06/19& 27 & Sgr 1 &  1800 \\
1998/06/19& 27 & Sgr 1 &  2400 \\
1998/06/19& 27 & Sgr 1 &  2400 \\
1998/06/19& 27 & Sgr 1 &  1380 \\
1998/06/19& 30 & Sgr 1 &  3600 \\
\tableline

\end{tabular}
\end{center}
\end{table}

\section{Radial velocities}

The spectra were calibrated in wavelength using a He-Ar lamp
spectrum and zero-point shifts were determined from
atmospheric emission lines.
To measure radial velocities we used 
cross-correlation technique in the 480-530 nm
range. As templates we used 
synthetic spectra of appropriate temperature and
[Fe/H]$=-0.5$ computed with the SYNTHE code (Kurucz 1993).
We selected as likely Sgr members all the stars
 with $\rm 100 kms^{-1} \le v_{hel}\le 180 kms^{-1}$ (Ibata et al 1997).
This left us with a sample of 23 stars out of 57 observed. 
The velocity dispersion
of these stars is 23 kms$^{-1}$. Since the internal velocity
dispersion of Sagittarius is only 10 kms$^{-1}$ (Ibata et al 1995), we 
assume
the dispersion to be dominated by the measurement error of which
it may be taken as a somewhat conservative estimate.
The radial velocities are given in Table 2, the stars are identified
by their number in the Marconi et al (1998) paper. 

\section{Abundances }

\begin{figure}[t]
\plotone{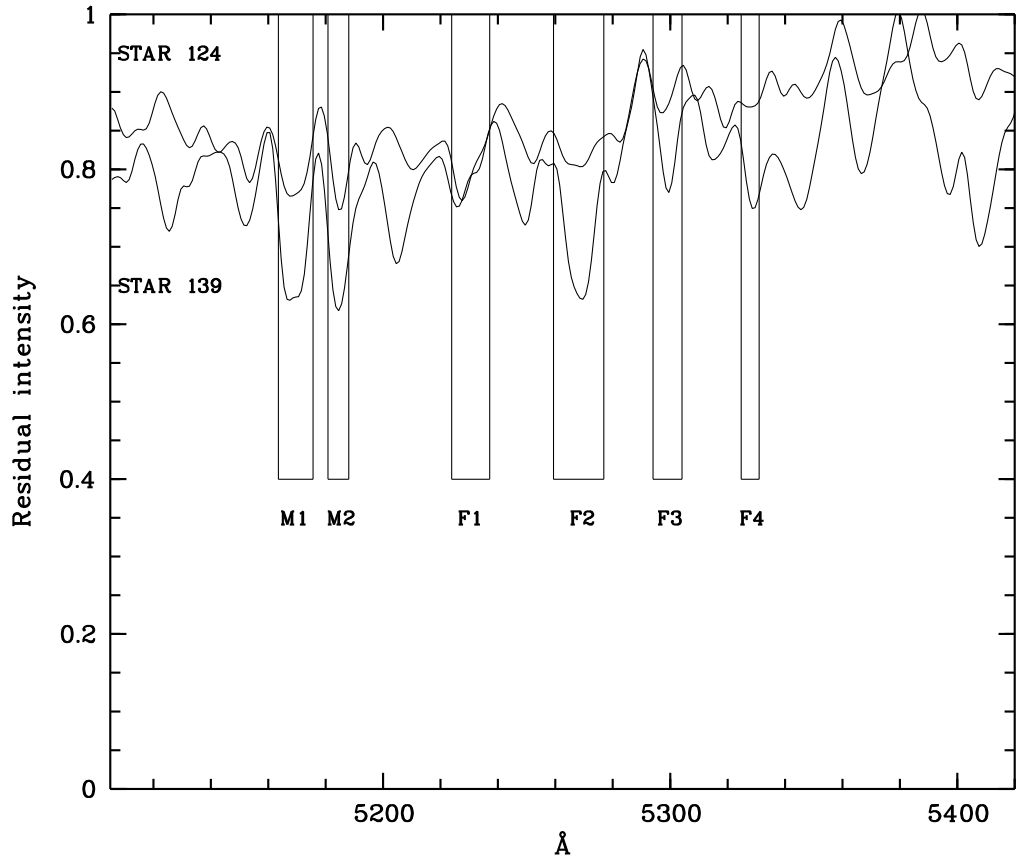}
\caption{
Comparison of the spectra of stars 124 and 139, which
have the same $(V-I)_0$. The difference in line strengths has
to be assigned to a different metallicity.
The spectra have been smoothed with a gaussian of
width 300 kms$^{-1}$ for display purposes.
The solid boxes show the spectrophotometric indices used. }
\end{figure}

For each star we determined the effective
temperature from the $(V-I)_0$ colour of Marconi et al (1998)
and the calibration of Alonso et al (1996).
Although the calibration is only valid for dwarfs,
$V-I$ depends little on gravity; 
based on  theoretical colours  computed by Castelli (1998)
we expect to make an error
less than 100 K, negligible in this context.
Gravities were determined from the  
 isochrones 
of Straniero, Chieffi \& Limongi (1997)
of 8 Gyr and [Fe/H]$=-0.5$.

We developed an iterative procedure for the estimate
of abundances.
For each star we started from a guess of the metallicity 
and computed a synthetic spectrum.
Then both the observed and computed spectra were 
pseudo-normalized using the same pseudo-continuum windows,
the observed spectrum was then normalized by multiplying
it by the ratio normalized/pseudo-normalized
obtained for the synthetic spectrum.
On this normalized spectrum we measured four
line indices, two of which measure the Mgb triplet
(M1 and M2)
and four which measure features due to iron-peak
elements (F1,F2,F3,F4). Placement of the indices
is illustrated in figure 1. The indices were defined so as to work
efficiently on the sun.
Then for each index we iteratively computed its value
on the synthetic spectrum, changing the input
abundance until it matched the measured index.
We first iterated only on the F (iron-peak) indices
at each iteration scaling all of the abundances of 
the input model. Next we took as scaling the median
of the four F indices and keeping this value fixed
we iterated on the M indices changing only the Mg abundance.
Next we computed a new synthetic spectrum using
this scaling and Mg abundance and re-normalized
the observed spectrum on this new synthetic spectrum
and iterated the whole procedure until the change in
metallicity was less than 0.2 dex. Typically three
iterations were enough to meet this criterion.
The procedure is quite CPU intensive and each iteration takes about
three hours on the  DEC Alpha of the OAT. 

The effectiveness of the procedure was tested
on the solar spectrum using both the Kurucz et al (1984)
solar atlas, degraded at the resolution of our grism
spectra and twilight spectra.
We were able to reproduce the solar (Fe/H) and (Mg/Fe)
to within 0.1 dex. The values determined
from the solar atlas for each index were used as zero point,
so as to make our analysis strictly differential 
with respect to the Sun.

Three matters are of main concern  
in assessing the results obtained in this way:
1) the reliability of the line list we are using. The
good results obtained for the Sun 
do not grant the completeness and
correctness of the list, since our program stars are about 1000 K
cooler and of much lower gravity.  We shall
try to quantify this issue in the future by analyzing the spectrum 
of a cool giant;
2) the placement of the continuum. The EMMI slitlets
have ragged edges which do not allow a flux calibration, 
this, in conjunction with the low S/N of our spectra
may induce an error  in the continuum which is difficult to asses;
3) the microturbulent velocity has to be assumed, the results
presented here are obtained for $\rm \xi = 2~ kms^{-1}$. Clearly at
this resolution we have no handle on microturbulence
what we shall do in a forthcoming paper is to quantify
the effect of microturbulence on abundances.
While these factors may affect our analysis in a systematic 
way (i.e. we may have a zero point uncertainty in the derived metallicities)
they will not affect the relative abundances of the analyzed
stars since the effective temperatures and gravities are quite similar.

\begin{table}[t]
\caption{Results}
\begin{center}
\begin{tabular}{ccccccc}
\tableline
\#&
$V_0$&
$v_r$&
$T_{eff}$&
$\rm log~ g$&
$\rm [Fe/H]\hfill$&
$\rm [Mg/Fe]\hfill$\\
 & & kms$^{-1}$& K &\\
\tableline
105 & 17.55 & 112 & 5041 & 2.59 & $-0.5$ & $+0.1$\\
115 & 17.84 & 144 & 4953 & 2.36 & $-1.0$ & $+1.2$\\
124 & 17.00 & 114 & 4891 & 2.21 & $-0.6$ & $+0.3$\\
128 & 17.40 & 145 & 4778 & 1.93 & $+0.7$ & $-0.5$\\
139 & 17.78 & 147 & 4891 & 2.21 & $-0.1$ & $+0.4$\\
141 & 17.45 & 161 & 4977 & 2.42 & $-0.6$ & $+0.4$\\
142 & 17.55 & 149 & 5118 & 2.81 & $-0.7$ & $+0.2$\\
201 & 17.54 & 135 & 5003 & 2.49 & $+0.1$ & $-0.0$\\
\tableline
\end{tabular}
\end{center}
\end{table}

\section{Discussion}

The results  for eight of the 23
candidate members of Sgr are presented in Table 2.
Two facts stand out: 1) we do not find any star as
metal--poor as [Fe/H]$=-1.5$, as was suggested by the photometry ;
2)  three out of eight stars are metal--rich.
Our metallicity scale may still suffer from a zero-point uncertainty
which could make the most metal--poor stars we observe
close to $-1.5$, it is however interesting to note
that a  mismatch between photometric and spectroscopic
metallicities in the same sense as the one we have here
has been found by Francois et al (1997)
for the young halo globular cluster Ruprecht 106.
Investigation of abundances of the whole available sample
might allow to establish if we are in the presence of
some systematic effect in abundance determinations or
if what we see is merely a selection effect.

The presence of the metal-rich stars is disturbing,
although they may in fact just be Bulge interlopers,
three out of eight seems a rather high percentage.
Furthermore the metal--rich stars do not show
sign of alpha--element enhancement as that displayed by
the Bulge K giants analyzed by McWilliam \& Rich (1994),
but are more like the SMR stars in the solar neighbourhood.
An independent investigation of metallicities
of Sgr giants has been carried out by Smecker-Hane, McWilliam
\& Ibata (1998) based on high-resolution spectra obtained
at the Keck telescope, their results indicate
a metallicity $\sim -1.5$ for the metal--poor stars, lower 
than our lowest metallicity stars,
however two out of seven stars they have analyzed are metal--rich
and alpha--elements O and Ca are underabundant with  respect
to solar, in keeping with our results.

Probably the best way to decide whether these metal-rich stars
belong  to the Bulge or to Sagittarius
is to determine the proper motions of the stars in their field.
Bulge stars should show  larger proper motions while Sagittarius
stars should show a small and coherent proper motion.
Such an investigation is within the capabilities of
a diffraction limited instrument, such as HST, over a time baseline
of just few years.
At the distance of 8.5 kpc a star with a transverse motion of 100
kms$^{-1}$ would show a proper motion of 2.48 mas/yr.

The future for abundance determinations for Sagittarius
lies in the use of 8m class telescopes.
The 8.2m VLT with the FORS1 instrument
will allow to obtain spectra of these stars with S/N $\sim 100$, albeit
with a slightly lower resolution. Moreover the slitlets
of FORS1 in multi-object mode are sharp and will allow 
flux calibration of
the spectra. Our technique of abundance estimation 
applied to such spectra should
then be capable of producing abundances accurate to the 0.3 dex
level.

\acknowledgments

We wish thank F. Castelli for providing the theoretical
colours and constant support in using
the model--atmosphere and spectrum synthesis codes
and A. Chieffi for providing the isochrones
used in this work.
Part of this work was done while P.B.
was at ESO-Garching
as a visiting scientist.

\end{document}